\documentclass[prl,floatfix,showpacs,amsmath,twocolumn]{revtex4}
\usepackage{amssymb}
\usepackage{graphics}
\usepackage{epsfig}
\def\qed{\leavevmode\unskip\penalty9999 \hbox{}\nobreak\hfill
     \quad\hbox{\leavevmode  \hbox to.77778em{%
               \hfil\vrule   \vbox to.675em%
               {\hrule width.6em\vfil\hrule}\vrule\hfil}}
     \par\vskip3pt}
\def\ibb #1{\leavevmode\hbox{\kern.3em\vrule
     height 1.5ex depth -.1ex width .4pt\kern-.3em\rm#1}}

\newcommand{\be}{\begin{equation}}
\newcommand{\ee}{\end{equation}}
\newcommand{\bea}{\begin{eqnarray}}
\newcommand{\eea}{\end{eqnarray}}
\newcommand{\la}{\langle}
\newcommand{\ra}{\rangle}

\begin{document}

\title{Entanglement versus Correlations in Spin Systems}
\author{F. Verstraete, M. Popp and J.I. Cirac}
\address{ Max-Planck-Institut f{\"u}r Quantenoptik,
         85748 Garching, Germany}
\date{\today}

\date{\today}

\begin{abstract}
We consider pure quantum states of $N\gg 1$ spins or qubits and
study the average entanglement that can be \emph{localized}
between two separated spins by performing local measurements on
the other individual spins. We show that all classical correlation
functions provide lower bounds to this \emph{localizable
entanglement}, which follows from the observation that classical
correlations can always be increased by doing appropriate local
measurements on the other qubits.  We analyze the localizable
entanglement in familiar spin systems and illustrate the results
on the hand of the Ising spin model, in which we observe
characteristic features for a quantum phase transition such as a
diverging entanglement length.
\end{abstract}

\maketitle

The mathematical description of multiparticle quantum systems
plays an important role in several branches of physics. The main
difficulty stems from the fact that the number of parameters
needed to describe a quantum state grows exponentially with the
number of particles. However, sometimes it is possible to capture
the most relevant physical properties by describing these systems
in terms of very few parameters. This is the case, for example, in
quantum statistics, where two--particle correlations play a
fundamental role. They allow us to understand several complex
physical phenomena, like phase transitions. Furthermore, they give
rise to concepts like correlation length, which quantifies a very
intuitive property of these systems.

Multiparticle systems are also of central interest in the field of
quantum information and, in particular, the quantification of the
entanglement contained in quantum states. The reason is that
entanglement is {\it the} physical resource to perform some of the
most important quantum information tasks, like quantum information
transfer (cfr. teleportation) or quantum computation.

Given the common interest of quantum statistical mechanics and
quantum information in multiparticle systems it is natural to try
to describe the physical phenomena, like quantum phase
transitions, appearing in (e.g.) spin systems from the point of
view of entanglement. The main restriction one encounters is the
fact that there exist very few measures of multiparticle
entanglement with a clear physical meaning. In any case, since
entanglement measures correlations (for pure states), one would
expect that reasonable entanglement measures be intimately
connected to the correlation functions widely used in the context
of quantum statistical mechanics. This is not the case, however,
if one studies the behavior of the entanglement of formation
between two separate spins after tracing out the rest
\cite{O02,N02,verz1,verz2,verz3}. Although this approach exhibits
a very pronounced (universal) behavior of this quantity at the
transition point, it rapidly vanishes as the distance between the
spins goes beyond 3 or 4, and thus it is not related to the
correlations possessed by the state. Note also that the approach
by Vidal et al. \cite{V02}, in which they study the scaling of
entanglement of a block of spins with the other ones, does not
quantify the entanglement as a resource (that could be used, for
example, for teleportation).

In this Letter we introduce a new concept which we call
Localizable Entanglement (LE). On the one hand, this quantity has
a very well defined physical meaning which treats entanglement as
a truly physical resource. On the other, it establishes a very
close connection between entanglement and correlation functions,
as one would naturally expect. The LE of two particles is the
maximal amount of entanglement that can be localized in these two
particles, on average, by doing local measurements in the rest of
the particles. The LE naturally leads to the definition of
\emph{entanglement length}, which measures the typical length
scale at which the LE decays. The LE has an operational meaning
which applies to situations in which out of some multiparticle
entangled state one would like to concentrate as much entanglement
as possible in two particular particles. This occurs, for example,
in the context of quantum repeaters \cite{Br98} or in the context
of quantum transport with spin systems (spintronics \cite{loss}).
The determination of the LE is a formidable task since it involves
an optimization over all possible local measurement strategies,
and thus cannot be determined in general. We have nevertheless
managed to determine tight upper and lower bounds. The first ones
stem from considering joint measurements on the rest of the spins,
which is closely related to the concept of entanglement of
assistance \cite{DiV98}. The second and more interesting ones can
be derived by proving that there always exist local measurement
which do not decrease the existing correlations between the two
spins, something that despite its generality, up to our knowledge
has not been considered in the context of quantum information.

The usefulness of these findings will be illustrated on the hand
of the entanglement present in the ground states of standard spin
Hamiltonians. Let us however first consider a simple example
involving the $N$-qubit GHZ-state:
\[|GHZ\rangle=\frac{1}{\sqrt{N}}\left(|00\cdots 0\rangle+|11\cdots
1\rangle\right).\] In this case, the LE is maximal as  it is
possible to create a Bell-state between two arbitrary qubits by
measuring the other qubits in the $|+\rangle,|-\rangle$ basis
(here $|\pm\rangle=(|0\rangle\pm|1\rangle)/\sqrt{2}$). The
existence of these quantum correlations could also have been
revealed by studying the classical correlation functions
\[Q_{\alpha\beta}^{ij}=\langle\psi|\sigma_\alpha^i\otimes\sigma_\beta^j|\psi\rangle-\langle\psi|\sigma_\alpha^i|\psi\rangle\langle\psi|\sigma_\beta^j|\psi\rangle,\]
where $i,j$ denote the positions of the spins under interest and
$\alpha,\beta$ label the Pauli matrices. The correlation in the
$\alpha=\beta=z$-direction is the maximal possible one. It will
indeed be shown that correlation functions yield lower bounds to
the LE. However, in general the presence or absence of classical
correlations only gives a coarse-grained picture of the
entanglement that ought to be created between two distant spins.
It is easy to find examples of highly entangled quantum states
exhibiting no classical correlations whatsoever between any pair
of spins. As an example, consider the so-called cluster states
\cite{Br01}, obtained by the unitary evolution of an initially
separable state under the action of the Ising Hamiltonian:
\[|\psi\rangle=\frac{1}{2^{N/2}}\left(\otimes_{i=1}^{N-1}(|0\rangle_i\sigma_z^{(i+1)}+|1\rangle_i)\right)(|0\rangle_N+|1\rangle_N)\]
When $N\geq 5$, all reduced 2-qubit density operators are
proportional to the identity and hence no correlations exist
between any two spins. However, suitable local measurements on any
$N-2$ qubits can always create a Bell-state between the two
remaining ones \cite{Br01}, hence indicating maximal LE.

Let us next give a formal definition of the LE. Consider a pure
state $|\psi\rangle$ of $N$ spins. Then the localizable
entanglement $E_{ij}(\psi)$ is variationally defined as the
maximal amount of entanglement that can be created (i.e.
localized), on average, between the spins $i$ and $j$ by
performing local measurements on the other spins. More
specifically, every measurement basis specifies a pure state
ensemble ${\mathcal{E}} = \{p_s , | \phi_s \ra \} $ consisting of
at least $2^{(N-2)}$ elements counted by the index $s$. In this
notation $p_s$ denotes the probability to obtain the two-spin
state $|\phi_s \ra$ after performing the measurement $|s\ra$ on
the assisting spins of our N partite spin system. The LE is then
defined as \[ E_{ij}= \max_{\mathcal{E}} \sum_s p_s \ E(|\phi_s
\ra) , \] where $E(|\phi_s\rangle)$ denotes the entanglement of
$|\phi_s\rangle$. As we deal with pure states of two qubits, all
entanglement measures are essentially equivalent, and in order to
make the connection with correlation functions, we will measure
the entanglement on the hand of the concurrence \cite{Woo98}:
indeed, it can readily be checked that the maximal correlation
function \footnote{The basis with maximal possible correlation can
in general be found by calculating the singular value
decomposition of the $3\times 3$ matrix $Q_{\alpha\beta}$, and the
largest singular value corresponds to the maximal possible
correlation.} for a pure state of two qubits coincides with the
concurrence, given by
$C(|\psi\rangle=a|00\rangle+b|01\rangle+c|10\rangle+d|11\rangle)=2|ad-bc|$.
Note that the behavior of the LE as defined in terms of the
entropy of entanglement would be very similar \footnote{For a
given concurrence $C$, the entropy of entanglement is given by
$f(C)=H\left((1+\sqrt{1-C^2})/2\right)$ with $H(x)$ the Shannon
entropy. Due to the convexity of $f(C)$, the following bounds
hold: $f\left(\sum_s p_s C_s\right)\leq \sum_s p_s f(C_s)\leq
\sum_s p_s C_s$.}.

Due to the variational definition, the LE is very difficult to
calculate in general. Moreover, in typically large spin systems
one does not have an explicit parametrization of the state under
interest, but just information about the classical 1- and
2-particle correlation functions (which parameterize completely
the 2-qubit reduced density operator). It would therefore be
interesting to derive tight upper and lower bounds to the LE
solely based on this information. The upper bound can readily be
obtained using the concept of entanglement of assistance
\cite{DiV98}, which would correspond to the LE if global or joint
measurements were allowed on the other spins. It was shown in
\cite{V01} that the entanglement of assistance can be calculated
as follows: given the reduced density operator $\rho_{ij}$ and a
square root $X$, $\rho_{ij}=XX^\dagger$, then the entanglement of
assistance $E_{ij}(\psi)$ as measured by the concurrence is equal
to the trace norm ${\rm Tr}|X^T(\sigma_y\otimes\sigma_y)X|$.

The lower bound is more subtle, and will be shown to follow from
the following interesting theorem: \emph{given a (pure or mixed)
state of $N$ qubits with classical correlation
$Q^{ij}_{\alpha\beta}$ between the spins $i$ and $j$ and
directions $\alpha,\beta$, then there always exist directions in
which one can measure the other spins such that this correlation
do not decrease, on average.} This automatically implies that
there always exist local measurements that increase (or keep) the
classical correlations. Surprisingly, this very general theorem
seems not to have been noticed before, and is interesting on its
own. It could be very useful in the context of cryptography (e.g.
multipartite distribution of common randomness \cite{Dev03}).

Let us next proof this theorem. Note that it is sufficient to
consider mixed states of three qubits. Let us parameterize a mixed
3-qubit density operator by four $4\times 4$ blocks
\[\rho=\left[\begin{array}{cc} \rho_1 &\sigma\\\sigma^\dagger
&\rho_2\end{array}\right].\] Without loss of generality, let us
consider the $Q_{zz}^{12}$ correlations. The original correlations
are completely determined by the diagonal elements of the density
operator $\rho_1+\rho_2$. A von Neumann measurement in the
$|\pm\rangle:=\cos(\theta/2)|0\rangle \pm\sin(\theta/2)\exp(\pm
i\phi)|1\rangle$ basis on the third qubit results in the hermitian
unnormalized 2-qubit operators
\begin{eqnarray*}
X_\pm&=&\frac{\rho_1+\rho_2}{2}\pm\cos(\theta)\frac{\rho_1-\rho_2}{2}\\
&&\hspace{.3cm}\pm\sin(\theta)\left(\cos(\phi)\frac{\sigma+\sigma^\dagger}{2}+\sin(\phi)\frac{i(\sigma-\sigma^\dagger)}{2}\right).\end{eqnarray*}
Defining $p_\pm={\rm Tr}(X_\pm)$, we have to prove that there
always exist parameters $\theta,\phi$ such that the correlation,
on average, does not decrease:
\[p_+|Q_{zz}(X_+/p_+)|+p_-|Q_{zz}(X_-/p_-)|\geq|Q_{zz}(X_++X_-)|.\]

Without loss of generality we can assume that
$\alpha=Q_{zz}(X_++X_-)$ is positive. Removing the absolute value
signs and parameterizing
\[\bar{x}:=[\cos(\theta);\sin(\theta)\cos(\phi);\sin(\theta)\sin(\phi)],\]
some straightforward algebra yields the sufficient inequality
\begin{equation}
\bar{x}^T\left[\alpha\left(\bar{c}-\frac{\bar{\beta}}{\alpha}\right)\left(\bar{c}-\frac{\bar{\beta}}{\alpha}\right)^T+Q-\frac{\bar{\beta}\bar{\beta}^T}{\alpha}\right]x\geq
0\label{ineq2}\end{equation} where $\bar{c}$ is such that
$p_{\pm}=(1\pm\bar{c}^T\bar{x})/2$, and $\bar{\beta},Q$ are
defined as $3\times 1$ and $3\times 3$ blocks of the matrix
\[S=R^T(\sigma_y\otimes\sigma_y)R=\left[\begin{array}{cc} \alpha
&\bar{\beta}^T\\
\bar{\beta} & Q\end{array}\right].\] Here $R$ is the real $4\times
4$ matrix whose columns consist of the diagonal elements of the
matrices $(\rho_1+\rho_2)$, $(\rho_1-\rho_2)$,
$(\sigma+\sigma^\dagger)$, $i(\sigma-\sigma^\dagger)$.

Due to Sylvester's law of inertia \cite{Horn85}, we know that $S$
has two positive and two negative eigenvalues. Now
$Q-\bar{\beta}\bar{\beta}^T/\alpha$ is the Schur complement of
$\alpha$, and hence corresponds to a principal $3\times 3$ block
of the matrix $S^{-1}$. Due to the interlacing properties of
eigenvalues of principal blocks \cite{Horn85}, it follows that it
has either two positive and one negative eigenvalue or two
negative and one positive one. And this of course ensures that
there always exists an $\bar{x}$ such that the inequality
(\ref{ineq2}) is fulfilled, completing the proof.\qed

Note that the proof is constructive and allows to determine a
measurement strategy that would at least achieve the bound
reported. Note also that exactly the same proof applies when
correlations between larger blocks of spins would be considered
\cite{inprogress}.

Let us now show how this theorem yields a lower bound to the LE.
Given an initial pure state of $N$ qubits, we know that the
measurement of the first, second, ... $N-3$'th qubit can be chosen
such that on average the final correlations do not decrease. But
we end up with a pure state of two qubits, for which the
concurrence is equal to the maximal correlation. A lower bound to
the LE is therefore given by the maximal correlation function, and
following the previous theorem, there is always a constructive way
of determining a measurement strategy that achieves this lower
bound. Surprisingly, we will see that this lower bound seems to be
the exact value for the LE in the case of many systems of
interest.

The previous findings can readily be applied to the study of
entanglement in translational invariant ground and excited states
of spins arranged in a regular lattice. In quantum statistics and
more specifically in the study of quantum phase transitions, the
correlation length is the canonical parameter of interest. The
concept of LE readily lends itself to define the related
\emph{entanglement length} $\xi_{E}$ as the typical length scale
at which it is possible to create Bell states by doing local
measurements on the other spins. More specifically, the
entanglement length is finite if and only if the the LE
$E_{i,i+n}\simeq\exp(-n/\xi_E)$ decays exponentially in $n$, and
the entanglement length $\xi_E$ is defined as the constant in the
exponent in the limit of an infinite system (see also Aharonov
\cite{Ahar}):
\[\xi_{E}^{-1}=\lim_{n\rightarrow\infty}\left(\frac{-\log
E_{i,i+n}}{n}\right)\] Note that a diverging correlation length
automatically implies a diverging entanglement length and hence
long-range quantum correlations, although the converse is not
necessarily true\footnote{This leaves open the possibility of
Hamiltonians exhibiting a phase transition as indicated by a
diverging entanglement length at the critical point but with a
correlation length remaining finite.}.

More specifically, we have studied ground states of two spin
interaction Hamiltonians with parity symmetry
$\otimes_{i=1}^N\sigma_z^i$ of the form:
\[\mathcal{H}=- \sum_{i,j} \ \sum_{\alpha=x,y,z} \gamma_{\alpha}^{ij}
\sigma_\alpha^i \sigma_\alpha^j - \sum_i \gamma^i \sigma_z \ , \]
with coupling coefficients $\gamma_{x}^{ij}\geq\gamma_{y}^{ij}\geq
0$ (note that this model also  includes spin systems in higher
dimensional lattices). Extensive numerical calculations on systems
of up to 20 qubits showed that our lower bound is always very
close to the LE, and typically is even equal to the exact value of
the LE \cite{inprogress}: this is surprising and highlights the
power of the given lower bound. Note also that whenever the parity
symmetry is present, the upper and lower bounds can easily be
calculated in terms of the correlation functions:
\begin{eqnarray*}
\max\left(Q_{xx}^{ij},Q_{yy}^{ij},Q_{zz}^{ij}\right)&\leq&
E_{ij}\leq\frac{\sqrt{s_+^{ij}}+\sqrt{s_-^{ij}}}{2}\\
s_{\pm}^{ij}&=& \left( 1 \pm \la \sigma_z^i\sigma_z^j \ra
\right)^2- \left(\la \sigma_z^i \ra \pm \la \sigma_z^j
\ra\right)^2
\end{eqnarray*}

As an illustration, we consider the ground state of the Ising
Hamiltonian
($\gamma_\alpha^{ij}=\lambda\delta_{\alpha,x}\delta_{j,i+1};\gamma^i=1$),
which has been solved exactly \cite{P70} and exhibits a quantum
phase transition at $\lambda=1$. In this case, the maximal
classical correlation is always given by $Q_{xx}$, and we
conjecture that it is equal to the LE (for more details, we refer
to \cite{inprogress}). For $\lambda<1$, the LE is small as the
ground state is almost separable, and the entanglement length is
finite (see Fig. 1) as the LE decreases exponentially with the
spin distance. At the quantum critical point $\lambda=1$, the
behavior of the LE changes drastically, as it decreases
polynomially $E_{i,i+n} \sim n^{-1/4}$, thus leading to a
diverging entanglement length $\xi_E$. In the case $\lambda > 1$
we also get $\xi_E=\infty$, as the LE saturates to a finite value
given by $M_x^2=1/4(1-\lambda^{-2})^{1/4}$. Indeed, the ground
state is then close to the GHZ-state. This behavior is illustrated
in Fig. 1.

\begin{figure}[h]
      \epsfig{file=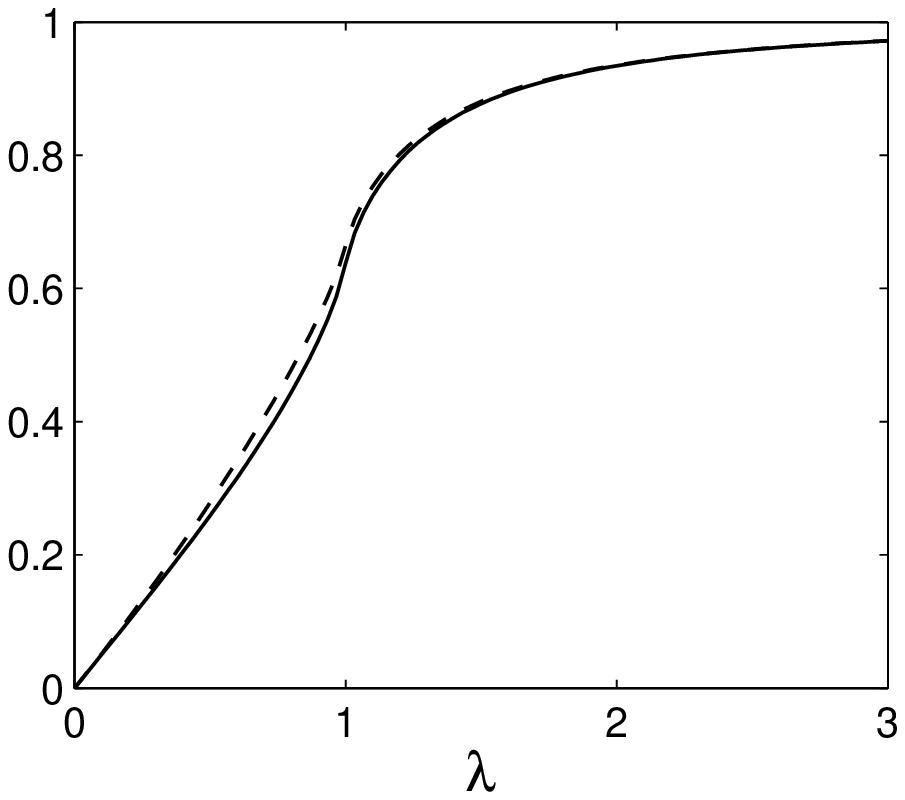,angle=-90,width=0.49\linewidth}
      \epsfig{file=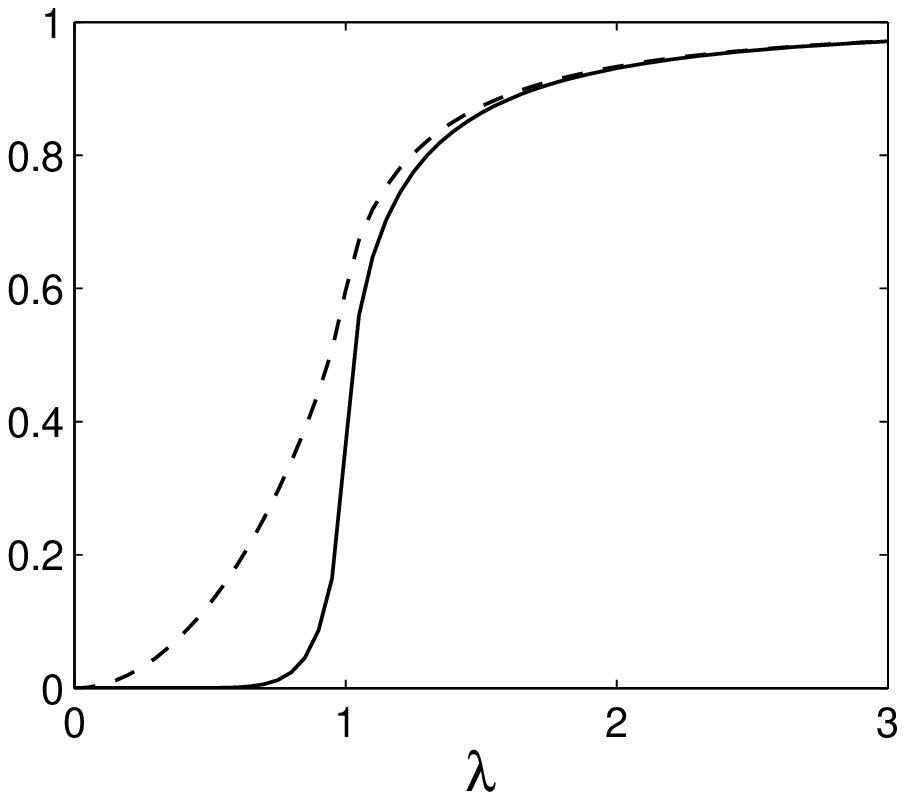,angle=-90,width=0.49\linewidth}
\caption{Localizable Entanglement $E_{ij}$ and Correlation
function $Q_{xx}^{ij}$ (solid) and upper bound (dashed) as a
function of the coupling parameter $\lambda$ for the infinite
Ising chain with spin distance $n=1$ (left) and $n=10$ (right).}
       \label{fig2}
\end{figure}

In a more realistic setup however, the parity symmetry of the
Ising Hamiltonian will be broken by a perturbation and the ground
state for large coupling will also be separable. Indeed, the
energy gap between the lowest energy states with different parity
decays exponentially in the number of qubits in the region
$\lambda>1$, and henceforth the ground state becomes a
superposition of these two states. The calculation of its LE and
the corresponding lower and upper bounds are depicted in Fig. 2.
\begin{figure}[t]
      \epsfig{file=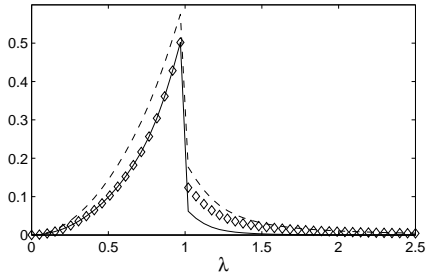,width=0.65\linewidth}
\caption{Correlation function $Q_{xx}^{i,i+4}$ (solid),
Localizable Entanglement $E_{i,i+4}$ (diamonds), and its upper
bound (dashed) as a function of the coupling parameter $\lambda$
for the ground state of the finite Ising chain ($N=14$) in the
case of broken parity symmetry by a small perturbating magnetic
field in the $x$-direction.}
       \label{fig-pert}
\end{figure}

These results give a clear illustration of the intimate connection
between classical correlations and entanglement in the case of
ground states of translational invariant Hamiltonians. The
plethora of results concerning classical correlation functions,
such as diverging correlation length at quantum phase transitions,
can now be interpreted from the perspective of quantum information
theory; one could argue that the status of classical correlations
has been lifted to the one of (useful) quantum correlations.

In conclusion, we have introduced the notion of localizable
entanglement.  It has a nice operational meaning as it quantifies
the amount of useful entanglement that can be created between two
spins by doing local measurements on all other spins. We proved
that classical correlation functions always provide lower bounds
to the LE, showing that the presence of classical correlations is
sufficient to be able to create Bell-like quantum correlations. As
a side-product, we proved that classical correlations in
multipartite mixed quantum states can always increase by doing
appropriate measurements on the other qubits. Finally, we
demonstrated the usefulness of these concepts in the context of
spin systems on a lattice, provided a natural definition of
entanglement length, and showed that it diverges at a quantum
phase transition. Further generalizations and applications will be
presented elsewhere \cite{inprogress}.

\acknowledgements We acknowledge interesting discussions with J.
Garcia Ripoll. This work was supported in part by the E.C.
(projects RESQ and QUPRODIS) and the Kompetenznetzwerk
"Quanteninformationsverarbeitung" der Bayerischen Staatsregierung.

\end{document}